# Closed Spaces in Cosmology

**Helio V. Fagundes**
Instituto de Física Teórica
Universidade Estadual Paulista
São Paulo, SP 01405-900, Brazil
E-mail: helio@ift.unesp.br

**ABSTRACT**

This paper deals with two aspects of relativistic cosmologies with closed (compact and boundless) spatial sections. These spacetimes are based on the theory of General Relativity, and admit a foliation into space sections $S(t)$, which are spacelike hypersurfaces satisfying the postulate of the closure of space: each $S(t)$ is a 3-dimensional closed Riemannian manifold. The discussed topics are:
(1) A comparison, previously obtained, between Thurston's geometries and Bianchi-Kantowski-Sachs metrics for such 3-manifolds is here clarified and developed.
(2) Some implications of global inhomogeneity for locally homogeneous 3-spaces of constant curvature are analyzed from an observational viewpoint.

## 1. INTRODUCTION

In this work relativistic cosmological models that admit *closed* – i. e., *compact and boundless* – spatial sections $S(t)$ are studied. Writing the spacetime metric as

$$ds^2 = -dt^2 + dl^2, \tag{1}$$

where

$$dl^2 = \gamma_{ab}\left(t, x^1, x^2, x^3\right) dx^a dx^b, \tag{2}$$

$S(t)$ is a Riemannian, 3-dimensional, oriented manifold, with metric $dl^2$. The orientability condition, which also holds for time lines, is required for the existence of spinor structure [1] in spacetime. (We shall return to this question in the last Section.) $S(t)$ is not necessarily homogeneous, but is locally homogeneous in a sense to be defined below, and has a closed global topology. These assumptions substantially modify the results obtained with the majority of spatially homogeneously cosmologies, which have open spatial topologies. In a previous study [2] the author established a correspondence between the spatially homogeneous metrics of Bianchi [3] and Kantowski-Sachs [4] on the one side, and the eight geometrical types of Thurston [5] on the other. In Section 2 below we develop this theme, and clarify some inaccuracies in Ref. 2.

The other topic here discussed is the subject of Section 3. Some theorems are there proved on the relation between sources of radiation in cosmic space and the possibility of effects related to the closure of space being observed, in particular the production of multiple images of some of those sources.

In Section 4 some remarks are made on the interfacing of these questions with quantum mechanical problems.[1]

## 2. THURSTON'S GEOMETRIC TYPES AND BIANCHI-KANTOWSKI-SACHS TYPES

In [2] a correspondence was established between the eight geometric types of Thurston [5] and the classification of spatially homogeneous metrics that has become standard in relativistic cosmology (see Refs. 6-8, for example): the Bianchi types and the exceptional Kantowski-Sachs type. We shall here refer to these metrics collectively as the BKS types. Thurston's research is deemed an important advance in the theory of 3-manifolds (cf. Ref. 9). He arrived at the important discoveries that (1) there are only eight basic homogeneous geometries (up to an equivalence class defined in Section 2.1 below), that can be supported by closed 3-manifolds; and (ii) if a closed 3-manifold of a given topology admits one of these geometric types, then this type is unique (cf. Ref. 9, Theorem 5.2). This second fact produces a partial classification of the topology of 3-manifolds through the Thurston geometries that can be assigned to them. The importance of this classification for cosmology is that, once we assume that cosmic 3-space is both closed and locally homogeneous, we know which classes of topology and geometry it can have. Combining this fact with the wealth of known BKS modes [8] we get a ready-made catalog of closed models and their topologies.

In [2] a few properties of Thurston's type were presented, such as some topological invariants and examples of spaces of these types which are found in cosmological applications. Here we will mainly develop mathematical questions not covered, or only succinctly mentioned, in that paper. In order to set these problems in context, we will make a summary of the Thurston and BKS classification schemes.

The spaces that interest us here are the *3-dimensional orientable, connected, complete Riemannian manifolds.* These properties will be tacitly presupposed in all spaces to be discussed below.

### 2.1. Thurston's classification for closed spaces

A metric on $M$ is called *locally homogeneous* if, for any two points $p, q \in M$ there are neighborhoods $(U, p)$ of $p$ and $(V, q)$ of $q$, and an isometry $\phi : (U, p) \to (V, q)$. In this case the space $M$ is said to be locally homogeneous. It turns out that, if $M$ is locally homogeneous, then its universal covering space $\widetilde{M}$ is *both* locally homogeneous and homogeneous with respect to the metric it inherits from *M* (cf. Ref. 9, p.402). This metric on $\widetilde{M}$ is *locally* identical to its parent metric on *M*, since the covering map $\pi : \widetilde{M} \to M$ is locally an isometry.

[1] Here we incorporate the corrections and comments of an Addendum to the paper (gr-qc/9805103), published in [36].

Let *X* be a simply connected Riemannian 3-manifold, homogeneous with respect to an orientation-preserving group *G* of isometries, such that the *stabilizer* $G_x$ (see below) at each point $x \in X$ is compact. The pair $(X, G)$ is a geometry in the sense of Klein (cf. Ref. 9, p. 403). We say with Thurston and Scott that a space *M* is a *geometric structure* modeled on the geometry $(X, G)$ if $M \cong \widetilde{M}/\Gamma$, i.e., *M* is isometric to the quotient space of *X* by a subgroup Γ of *G*, the action of Γ on *X* being discontinuous, discrete and without fixed points. It follows that $X = \widetilde{M}$, the universal covering space of *M,* and that *M* is locally homogeneous with respect to the metric of $(X, G)$. Let *G* = Isom(*X*) be the maximal group of orientation-preserving isometries of *X.* Thurston's eight geometric types (T-types) classify the geometries $(X, \mathrm{Isom}[X])$ on which geometric structures for closed, locally homogeneous spaces *M* can be modeled. In Ref. 2 we gave several examples of 3-spaces *M* modeled on Thurston's geometries $(X, \mathrm{Isom}[X])$, and of their applications to cosmology.

The eight geometries $(X, \mathrm{Isom}[X])$ are unique up to an equivalence relation defined in Ref. 9, p. 474: *$(X, G)$ and $(X', G')$ are equivalent if there is a diffeomorphism of X onto $X'$ that takes the action of G onto the action of $G'$*, which must be isomorphic to *G.* Table I lists the model spaces *X* and most groups Isom[*X*]. For more details see [5,9]. See Ref. 9, or our brief comments in Ref. 2, on the aspect of *Seifert fibration* for six of Thurston's eight types, and the important point that for seven of these types the allowed topologies are classified.

Another aspect of Thurston's classification method refers to the stabilizers or isotropy groups $G_x$ of each point $x \in X$: $G_x = \{g \in \mathrm{Isom}(X);\ gx = x\}$. If $I(G_x)$ is the branch of $G_x$ containing the unity, then (Ref. 9, p. 475) $I(G_x)$ is SO(3), SO(2), or the trivial group. In the first case Isom(*X*) has dimension six and we have the spaces of constant curvature T1 to T3. In the second case are types T4 to T7, with dim(Isom[*X*]) = 4, which possess a local rotational symmetry. Readers familiar with Ellis's work [10] will recognize the parallel of these results with his *locally rotational symmetrical* (LRS) homogeneous spaces. See Section 2.5 below.

On most of Thurston's types we have little to add here to what was said in [2]. But the metric for type T6, shown there without proof, was derived by us from properties indicated for this type in Ref. 5 (not in Ref. 9, as said in Ref. 2). Our derivation follows.

The metric space *X* for type T6 is $\widetilde{T}_1(H^2)$ which is the universal covering space of the fiber bundle $T_1(H^2)$ whose base space is the hyperbolic plane $H^2$, the fiber at

**Table I.** Model spaces of Thurston and their orientation preserving isometry groups (based on Refs. 5, 9).

| Type | $X$ | Isom[$X$] |
|---|---|---|
| T1 | $S^3$ | SO(4) |
| T2 | $E^3$ | $R^3 \times \mathrm{SO}(3)$ |
| T3 | $H^3$ | PSL(2,**C**) |
| T4 | $S^2 \times E^{1(\mathrm{a})}$ | $[G(S^2)\times G(E^1)]_+$ |
| T5 | $H^2 \times E^{1(\mathrm{a})}$ | $[G(H^2)\times G(E^1)]_+$ |
| T6 | $\widetilde{T}_1(H^2)^{(\mathrm{b})}$ | $R \times \widetilde{G}(H^2)$ |
| T7 | $\mathrm{Nil}^{(\mathrm{c})}$ | See note (c) |
| T8 | $\mathrm{Sol}^{(\mathrm{d})}$ | See note (d) |

(a) $S^2$ is the sphere and $G(S^2)$ is its group of isometries. Similarly for the Euclidean straight line $E^1$ and the hyperbolic plane $H^2$. The + subscript indicates the orientation-preserving subgroup of the product group.

(b) See Section 3.1. $\widetilde{G}(H^2)$ is the universal covering group of $G(H^2)$.

(c) Nil is the group of Heisenberg matrices $\begin{pmatrix} 1 & x & z \\ 0 & 1 & y \\ 0 & 0 & 1 \end{pmatrix}, x, y, z \in R$. Isom(Nil) is too complex to be succinctly defined here.

(d) Sol can be represented by $R^3$ with the multiplication law $(x, y, z)$ $(x, y, z)(x', y', z') = (x + e^{-z}x', y + e^{z}y', z + z')$. As with Nil, Isom(Sol) will not be described here.

$p \in H^2$ being $T_1(p) = \{\mathbf{u} \in T(p);\ \mathrm{length}(\mathbf{u}) = 1\}$ – the submanifold of tangent vectors of unit length. The metric on $H^2$ can be written as

$$dl_H{}^2 = dx^2 + \cosh^2 x\, dy^2. \tag{3}$$

Let $p = (x, y) \in H^2$, and $\mathbf{u} \in T_1(p)$. Since $\mathbf{u}\cdot\mathbf{u} = 1$ in the metric (3), we have $u^1 = \cos x,\ u^2 = \mathrm{sech}\, x \sin x,$ with $0 \le x \le 2\pi$. For a displacement $dx$ we have the total differential

$$Du^a = \left(\frac{\partial u^a}{\partial x^c} + \Gamma^a{}_{bc} u^b\right) dx^c + \frac{\partial u^a}{\partial z} dz, \tag{4}$$

where $a,b,c = 1-2$, $x^1 = x, x^2 = y$, and of course the Christoffel symbols are those for metric (3). Among these the non-null ones are $\Gamma^1{}_{22} = -\sinh x \cosh x$, $\Gamma^2{}_{12} = \Gamma^2{}_{21} = \tanh x$. Substituting into (4) we get

$$\begin{cases} Du^1 = -\sinh x \sin z\, dy - \sin z\, dz, \\ Du^2 = \tanh x \cos z\, dy + \operatorname{sech} x \cos z\, dz. \end{cases} \tag{5}$$

The length of $D\mathbf{u}$ will then be given by $dl_3{}^2 = (Du^1)^2 + \cosh^2 x (Du^2)^2$, which is the metric on the fiber $T_1(p)$. Therefore the metric on T6 is

$$\begin{aligned} d\lambda^2 &= dl_H{}^2 + dl_3{}^2 \\ &= dx^2 + \cosh^2 x\, dy^2 + \left(dz + \sinh x\, dy\right)^2. \end{aligned} \tag{6}$$

## 2.2. The classification of Bianchi for homogeneous spaces

The original work of Bianchi [3] has been reorganized into a modern formalism by theoretical cosmologists. Here we shall generally adopt the scheme of MacCallum [7].

Given a Bianchi algebra, we shall call its corresponding maximal Lie group a *Bianchi group*. We use the invariant bases $\{\omega^a\}$ of 1-forms to characterize the nine Bianchi types. They are given in Table II, based on Ref. 8, Table 8.2; but for type BVI($A$) we have changed the *y*, *z* coordinates of Kramer et al. [8]. Calling the latter $y', z'$, the new coordinates are $y = (y' + z')/\sqrt{2}$, $z = (-y' + z')/\sqrt{2}$. Similarly, the new 1-forms are $\omega^2 = (\omega'^2 + \omega'^3)/\sqrt{2}$, $\omega^3 = (-\omega'^2 + \omega'^3)/\sqrt{2}$.

Let $\widetilde{M}$ be a 3-dimensional, homogeneous with respect to a Bianchi group *G*. Then the most general Riemannian metric, invariant under the action of *G*, which can be supported by $\widetilde{M}$ has the form

$$dl^2 = \gamma_{ab}\omega^a\omega^b, \tag{7}$$

where the coefficients $\gamma_{ab} = \gamma_{ba}$ are constants compatible with a Riemannian metric. It follows that *G* is a group of isometries of $\widetilde{M}$, which may be called a *Bianchi space.*

**Table II.** List of 1-forms which constitute invariant bases under the action of Bianchi groups (based on Ref. 8). Type BIII coincides with type BVI(1).

| Bianchi type | $\omega^1$ | $\omega^2$ | $\omega^3$ |
|---|---|---|---|
| BI | $dx$ | $dy$ | $dz$ |
| BII | $dx - z\,dy$ | $dy$ | $dz$ |
| BIV | $dx$ | $e^x dy$ | $e^x(dz + x\,dy)$ |
| BV | $dx$ | $e^x dy$ | $e^x dz$ |
| BVI($A$), $0 \le A \le 1$ | $dx$ | $e^{(A-1)x} dy$ | $e^{(A+1)x} dz$ |
| BVII($A$), $0 \le A \le 1$ | $dx$ | $e^{Ax}(\cos x\,dy - \sin x\,dz)$ | $e^{Ax}(\sin x\,dy + \cos x\,dz)$ |
| BVIII | $\cosh y \cos z\,dx - \sin z\,dy$ | $\cosh y \sin z\,dx + \cos z\,dy$ | $dz + \sinh y\,dx$ |
| BIX | $\cos y \cos z\,dx - \sin z\,dy$ | $\cos y \sin z\,dx + \cos z\,dy$ | $dz - \sin y\,dx$ |

We call the *reference metric*[2] of a given Bianchi type the expression (7) for $\gamma_{ab} = \delta_{ab}$, that is,

$$dl^2 = (\omega^1)^2 + (\omega^2)^2 + (\omega^3)^2 . \tag{8}$$

These reference metrics will be the instrument for our comparison between the types of Bianchi groups and Thurston's geometric types (see Section 2.5 below).

## 2.3. The exceptional metric of Kantowski and Sachs

Bianchi's nine types classify, with one exception, all homogeneous 3-dimensional metrics. The exceptional case is Kantowski and Sachs's metric

[2] This was called a *standard metric* in the published version of this paper.

$$dl^2 = a^2 dx^2 + b^2 (dy^2 + \sin^2 y\, dz^2)\,, \tag{9}$$

where $a^2, b^2$ are positive constants [4]. We call *Kantowski-Sachs* (KS) *space* a simply connected space with the topology of $R \times S^2$, where *R* is the straight line and $S^2$ is the sphere, whose metric has the form (9). Taking $a^2 = b^2 = 1$, we get the reference metric for this type of homogeneous 3-manifold:

$$dl^2 = dx^2 + dy^2 + \sin^2 y\, dz^2\,. \tag{10}$$

## 2.4. Results applying to both Bianchi and Kantowski-Sachs types

Closed 3-dimensional spaces *M* need not be homogeneous (see Section 3 below). Because of this we have to handle BKS metrics on them indirectly, through their universal covering spaces $\widetilde{M}$, which are always homogeneous. The next two theorems address this question.

*Theorem 2.1*. If a closed 3-space *M* admits a BKS metric, then *M* is locally homogeneous with respect to this metric.
*Prof*. The universal covering space $\widetilde{M}$ of $M$ is locally homogeneous with respect to the given BKS metric, because it is homogeneous with respect to the group *G* of the same BKS type. Therefore, given two points $p, q \in M$, if $\widetilde{p} \in \pi^{-1}(p)$, $\widetilde{q} \in \pi^{-1}(q)$, there are neighborhoods $(\widetilde{U}, \widetilde{p})$ of $\widetilde{p}$, $(\widetilde{V}, \widetilde{q})$ of $\widetilde{q}$ in $\widetilde{M}$, and $g \in G$, such that $g(\widetilde{U}, \widetilde{p})$ = $(\widetilde{V}, \widetilde{q})$. On the other hand, $\pi$ is locally an isometry, that is (choosing new neighborhoods $(\widetilde{U}, \widetilde{p})$, $(\widetilde{V}, \widetilde{q})$ if necessary), the restrictions $\phi$ = $\pi_{|(\widetilde{U},\widetilde{p})}$, $\psi$ = $\pi_{|(\widetilde{V},\widetilde{q})}$ are diffeomorphisms that conserve the metric. It follows that $(V, q) = \psi(\widetilde{V}, \widetilde{q}) = \psi g(\widetilde{U}, \widetilde{p}) = \psi g \phi^{-1}(U, p)$, where $\psi g \phi^{-1}$ is an isometry of the same BKS type as *g*, since the maps $\phi$ and $\psi$ may be seen as mere coordinate transformations. ■

Any closed, locally homogeneous 3-space *M* is diffeomorphic to the quotient space $\widetilde{M}/\Gamma$, where Γ is a discrete subgroup of Isom($\widetilde{M}$) [36,37]. Therefore, from the discussion in the beginning of Section 2.1, *M* possesses a geometric structure modeled on $(\widetilde{M}, \mathrm{Isom}[\widetilde{M}])$. Combining this result with Theorem 2.1, we see that any closed 3-space $M$ which admits a BKS metric is modeled on one of Thurston's geometric types.

Bianchi types are divided into classes A and B – cf. Ref. 7, for example. As noted by Koike et al. [37], if a closed 3-space $M$ has a Bianchi group *G* of either class A type VIII or any class B type, then the discrete group Γ in $M \cong \widetilde{M}/\Gamma$ cannot be a subgroup of *G*. But then $\dim(\mathrm{Isom}[\widetilde{M}]) > 3$. This explains why there are no closed spaces of class B types IV and VI($0<A<1$): their full isometry groups are 3-dimensional.

On the other hand, the remaining Bianchi types – which are BI, BII, BVI(0), BVII(0), and BIX – plus the KS-type, may admit compact quotients $M \cong \widetilde{M}/\Gamma$, where $\Gamma \subset G \subset \mathrm{Isom}(\widetilde{M})$,[3] with metrics different from their reference metrics. For then the action of $\Gamma$ leaves the basis $\{\omega^a\}$ in Eq. (7) invariant, and similarly for Eq. (9). In general such metrics have *variable* matrices **K**. It would be interesting to know the relation of these metrics to the totality of Riemannian metrics that can be assigned to a given T-type topology.

**Table III.** List of Kantowski-Sachs and Bianchi reference metrics and their principal sectional curvatures (based on Ref. 2).

| BKS type | $d\lambda^2$ | $K_1$ | $K_2$ | $K_3$ |
|---|---|---|---|---|
| KS | $dx^2 + dy^2 + \sin^2 y\, dz^2$ | $1$ | $0$ | $0$ |
| BI | $dx^2 + dy^2 + dz^2$ | $0$ | $0$ | $0$ |
| BII | $(dz - x\,dy)^2 + dy^2 + dz^2$ | $-3/4$ | $1/4$ | $1/4$ |
| BIV | $dx^2 + e^{2x}dy^2 + e^{2x}(dz + x\,dy)^2$ | $-3/4$ | $(-5 \pm 2\sqrt{5})/4$ | |
| BV | $dx^2 + e^{2x}(dy^2 + dz^2)$ | $-1$ | $-1$ | $-1$ |
| BVI($A$), $0 \le A \le 1$ | $dx^2 + e^{2(A-1)x}dy^2 + e^{2(A+1)x}dz^2$ | $1 - A^2$ | $-(1 \pm A)^2$ | |
| BVII($A$), $0 \le A \le 1$ | $dx^2 + e^{2Ax}(dy^2 + dz^2)$ | $-A^2$ | $-A^2$ | $-A^2$ |
| BVIII | $\cosh^2 y\, dx^2 + dy^2 + (dz + \sinh y\, dx)^2$ | $-1/4$ | $-1/4$ | $-5/4$ |
| BIX | $\cos^2 y\, dx^2 + dy^2 + (dz - \sin y\, dx)^2$ | $1$ | $1$ | $1$ |

The following theorem is needed for the comparison of the classifications in Section 2.5.

*Theorem 2.2.* If a closed 3-space admits a metric of a given BKS type, then it also admits the reference metric for this type.

[3] In the published version of this paper it was erroneously stated that this property held for all BKS types. This fact also led to the substitution of a weaker form of Theorem 2.2 here.

*Proof.* If $M$ admits one of the metrics (7) or (9), by Theorem 2.1 $M$ is locally homogeneous. Therefore (cf. Section 2.1) its universal covering space $\widetilde{M}$ is homogeneous with respect to the BKS group of this type. Hence $\widetilde{M}$ is a BKS space that admits any metric of form (7) or (9) for this type, in particular its reference metric.. But then $M \cong \widetilde{M}/\Gamma$, where $\Gamma \subset \mathrm{Isom}(\widetilde{M})$, also supports this reference metric. ■

In cosmology the spacetime metrics with BKS space sections $S(t)$ are of the form (1), with $dl^2$ given by (7) with $\gamma_{ab} = \gamma_{ab}(t)$ or by (9) with $a^2 = a^2(t)$, $b^2 = b^2(t)$.

A remark on the calculation of principal sectional curvatures $K_p$, $p$ = 1-3, of the BKS reference metrics, the result of which was published in Ref. 2. From the curvature tensor ${}^{(3)}R_{abcd}$, we get $K_p$ by Cartan's method (Ref. 11, Sections 169, 170): the *matrix of curvature* is $\mathbf{K} = (K_{AB})$ where $K_{AB} = K_{BA} = {}^{(3)}R_{abcd}$, $(Aab)$ and $(Bcd)$ being cyclic permutations of (123); the principal sectional curvatures are the eigenvalues $K_p$ ($p$ = 1-3) of $\mathbf{K}$, which appear in Table III.

**Table IV.** The correspondence between Thurston and BKS types.

| T-type | BKS-types |
|---|---|
| T1 | BIX |
| T2 | BI, BVII(0) |
| T3 | BV, BVII$(0 < A \leq 1)$ |
| T4 | KS |
| T5 | BVI(1) = BIII |
| T6 | BVIII |
| T7 | BII |
| T8 | BVI(0) |

## 2.5. Comparison between Thurston and BKS types

Table IV is a list of Thurston types and corresponding BKS types. This correspondence is *not* one-to-one. Types BI and BVII(0) correspond to the single type T2. The reason is that the 3-dimensional groups BI and BVII(0) are different transitive subgroups of the 6-dimensional group $\mathrm{Isom}(E^3)$, where $E^3$ is Euclidean space, which defines the geometry of T2. On the other hand, if we have an additional LRS symmetry (cf. Section 2.1) in a BVII(0) metric – say, if $d\lambda^2 = \gamma_{11}dx^2 + \gamma_{22}(dy^2 + dz^2)$ – then this metric is also of type BI. Similarly, for groups BV and BVII($A$), $A > 0$, with respect to hyperbolic space $H^3$, which defines T3 geometry; for example, the LRS metric of type BVII($A$), $d\lambda^2 = \gamma_{11}+ \gamma_{22}(dy^2 + dz^2)$, $A > 0$, can be expressed as a BV metric by the transformation $x \rightarrow x/A$. See also Ref. 7, p. 549, and Ref. 12.

The comparison between Thurston and BKS types was obtained as follows:

For types T1, T2, T3, the model spaces $X$ are the spaces of constant curvature $K$, for which dim(Isom[$X$]) = 6. The correspondence is then established with the reference BKS metrics $d\lambda^2$ of constant curvature $K$, according to Table III, and using Theorem 2.2. Thus T1 ↔ BIX for $K = 0$; T2 ↔ BI and BVII(0) for $K = 0$; and T3 ↔ BV and BVII($A$), $0 < A \leq 1$, for $K < 0$.

For the other types the problem is reduced to finding a coordinate system for $X$, space modeled in ($X$, Isom[$X$]) according to Thurston, if the metric on $X$ coincides with such that its metric has the same form as one of the BKS metrics. For, given a closed BKS reference metric $d\lambda^2$ by Theorem 2.2 $X$ admits the general metric of this type – Eq. (7) for Bianchi types or Eq. (9) for type KS. In cases T4 and T5 the structure of the model space $X$ allows us to infer their metric, which may be written in the forms $d\lambda^2$(KS) and $d\lambda^2$(BVI[1]), respectively. For T6 the metric is given by the expression (6) calculated above, which coincides with $d\lambda^2$(BVIII). Finally, for T7 and T8 Scott [9] writes down the metrics, which are equal to $d\lambda^2$ for types BII and BVI(0), respectively.

As explained in Section 2.4, types BIV and BVI($A$), $0 < A < 1$, stay out of this correspondence. This means that we cannot have *spatially closed* cosmologies with metrics of these types. As we see from Table III, these are the only BKS types whose reference metrics have the three principal curvatures $K_p$ different from each other. This was taken in Ref. 2 as an empirical rule to decide which BKS metrics cannot be assigned to closed spaces. On the other hand, Ellis and Schreiber [13] remark that closed spaces cannot support metrics of these types because "horizon-crossings and whimper singularities occur." We suspect the existence of a link between these facts, but could not find it.

It was seen in Section 2.1 that $\dim(\text{Isom}[\widetilde{M}]) > 3$ for types T1 to T7, with $I(G_x)$ = SO(3) or SO(2). So a closed space $M$ which admits a BKS metric of any type except BVI(0), also admits LRS metrics of the same type. In the case of BVI(0) this is not possible, because the equivalence of $(\widetilde{M}, \text{Isom}[M])$ and ($X$, Isom[$X$]) for type T8 implies dim(Isom[$X$]) = dim(Isom[$X$]) = 3, leaving no room for an additional continuous symmetry in $\widetilde{M}$. The Bianchi types corresponding to T1-T3, T5-T7 in Table IV coincide with the list of LRS Bianchi spaces in Ref. 7, p. 549.

The study of individual closed manifolds modeled on Thurston's geometries is beyond the scope of this work. We intend to pursue this question later, but a few remarks can now be made. For types T1 and T2 the closed orientable spaces are completely classified [11, 14]. For T3 the classification is still a subject of research, but a number of examples are known [15-17]. For types T4-T8 the Isom($X$) groups are not much larger than the corresponding BKS groups. One may try to combine these facts with the old-fashioned but thorough study of Bianchi [3], and with the known topological properties [1,2,5,9], to produce detailed pictorial maps of closed 3-spaces which admit BKS metrics.

## 3. LOCALLY HOMOGENEOUS BUT GLOBALLY INHOMOGENEOUS SPACES OF CONSTANT CURVATURE

### 3.1. Friedmann-Robertson-Walker (FRW) models with closed spatial sections

This section discusses some properties of cosmological models with FRW spacetime metrics and global topologies $R_+\times\Sigma$, where $R_+$ is the positive time semi-axis and $\Sigma$ is a closed 3-dimensional space. If we factor out the expansion factor $a^2(t)$ in the spatial part of the metric

$$ds^2 = -dt^2 + a^2(t)d\lambda^2, \tag{11}$$

then $d\lambda^2$ is the metric of constant curvature on the comoving spatial sections $\Sigma$.

From Ref. 14 we get the following characterization of a closed 3-space:

The complete connected Riemannian manifolds of constant curvature are known as *space forms*. Here we shall call *closed space* a compact 3-dimensional space form. In general a space form $\Sigma$ of curvature $K$ is isometric to $\tilde{\Sigma}/\Gamma$, where

$$\tilde{\Sigma} = \begin{cases} H^3, \text{ hyperbolic space, if } K < 0, \\ E^3, \text{ Euclidean space, if } K = 0, \\ S^3, \text{ spherical space, if } K > 0, \end{cases} \tag{12}$$

and $\Gamma$ is a group of isometries of $\tilde{\Sigma}$ acting freely and properly discontinuously (see Ref. 18, Theorem 6.5). $\Sigma$ is also isometric to the space $\overline{P}$ obtained from a *fundamental polyhedron P* with faces pairwise identified (cf. Ref. 19, p. 213, for example). Except for $S^3$, closed spaces have multiply connected topologies.

Closed spaces of null and positive curvatures have been completely classified (cf. Ref. 14, Sections 3.5 and 7.5). This is a high mathematical accomplishment, as yet barely explored by cosmologists. As for hyperbolic closed spaces ($K < 0$), they are still a subject of advanced research; see [5,15,17,20].

Up to now most cosmological studies involving FRW models of nontrivial space topology have restricted themselves to Einstein-de Sitter models with $\Sigma = T^3 = E^3/Z^3$, the flat 3-torus ($Z^3$ is the group generated by three independent, finite translations). See [21] and references there. But we have also obtained interesting results [16,22] from cosmologies with non-flat space sections of nontrivial topology.

### 3.2. Homogeneity and inhomogeneity of space forms

A space form Σ is locally homogeneous. According to the definition in Section 2.1, this means that, if $p, q \in \Sigma$, there are neighborhoods $(U, p)$ and $(V, q)$ which are related by an isometry. Since Einstein's equations are local, a solution of the form (11) that holds in $R_+ \times (U, p)$ will also hold in $R_+ \times (V, q)$.

The covering spaces $\tilde{\Sigma}$ are locally homogeneous and also *homogeneous*: their 6-dimensional groups of isometry $\tilde{G}$ act transitively on them. As for Σ, it may or may not be homogeneous (see Ref. 1, p. 11). In the latter case it is said to be *inhomogeneous* (or *globally inhomogeneous*, as in the title of this Section). Even if Σ is homogeneous, its isometry group is usually of smaller dimension than $\tilde{\Sigma}$, violating the full rotational symmetry of $\tilde{\Sigma}$. This fact led Hawking and Ellis [23] to reject such spaces as spatial sections of FRW models. But, as shown in Section 3.3 below, a reinterpretation of the cosmological principle will remove this objection. (Incidentally, some confusion may arise because for some physicists the idea of homogeneity is strongly associated with *uniformity of mass distribution*, which is called by Heller et al. [24] the *cosmological principle for the substratum*, in contrast to the *cosmological principle for the geometry*.)

*Definitions*. A *geodesic loop* $(x,\gamma)$, $x \in \Sigma \cong \tilde{\Sigma}/\Gamma$, $\gamma \in \Gamma$, is a geodesic segment $x(\lambda)$, $0 \le \lambda \le 1$, with $x(0) = x(1) = x$, that is lifted to segment $(\tilde{x}, \gamma\tilde{x})$ in Σ, where $\tilde{x} = l(x)$ is the point corresponding to $x$ in the fundamental polyhedron $P$ of Σ. (With this choice $\tilde{x}$ is uniquely defined on the fiber $\pi^{-1}(x)$, where $\pi$ is the covering map.) In particular, a *generator loop* $(x, \gamma_k)$ is a geodesic loop where $\gamma_k$ is one of the generators of Γ. The motivation for choosing geodesic loops is that they are the projections of light paths (null geodesics) of spacetime on the comoving spatial section Σ; this fact will be used in Theorem 3.3.

$D(x,z)$ will denote the distance between two points $x$, $z$, and $L(x,\gamma)$ the length of loop $(x,\gamma)$. Note that $L(x,\gamma) = D(\tilde{x}, \gamma\tilde{x})$.

The *loop angle* of $(x,\gamma)$ or angle$(x,\gamma)$, is the angle between the initial and final directions of $x(\lambda)$. If $\mathbf{h}$ is the metric on Σ, and $\mathbf{t}(\lambda)$ is the normalized tangent vector to the loop at $\lambda$, then angle$(x,\gamma)$ is the smallest positive determination of $\cos^{-1}\{\mathbf{h}[\mathbf{t}(0), \mathbf{t}(1)]\}$.

*Theorem 3.1*. If a space $\Sigma \cong \tilde{\Sigma}/\Gamma$ is homogeneous, then for any $x$, $x' \in \Sigma$, $\gamma \in \Gamma$, $L(x',\gamma) = L(x,\gamma)$.

*Proof.* From Ref. 14, Theorem 7.6.7, $\Sigma$ is homogeneous if and only if, for any $\tilde{x},\tilde{y} \in \tilde{\Sigma}$, $\gamma \in \Gamma$, we have $D(\tilde{x},\gamma\tilde{x}) = D(\tilde{y},\gamma\tilde{y})$. Now let $\tilde{x} = l(x)$, $\tilde{x}' = l(x')$. If $\Sigma$ is homogeneous, $L(x,\gamma) = D(\tilde{x},\gamma\tilde{x}) = D(\tilde{x}',\gamma\tilde{x}') = L(x',\gamma)$. ■

An example of this property is given in Fig. 1 in terms of the torus $\Sigma = T^2 \cong E^2/\Gamma$,

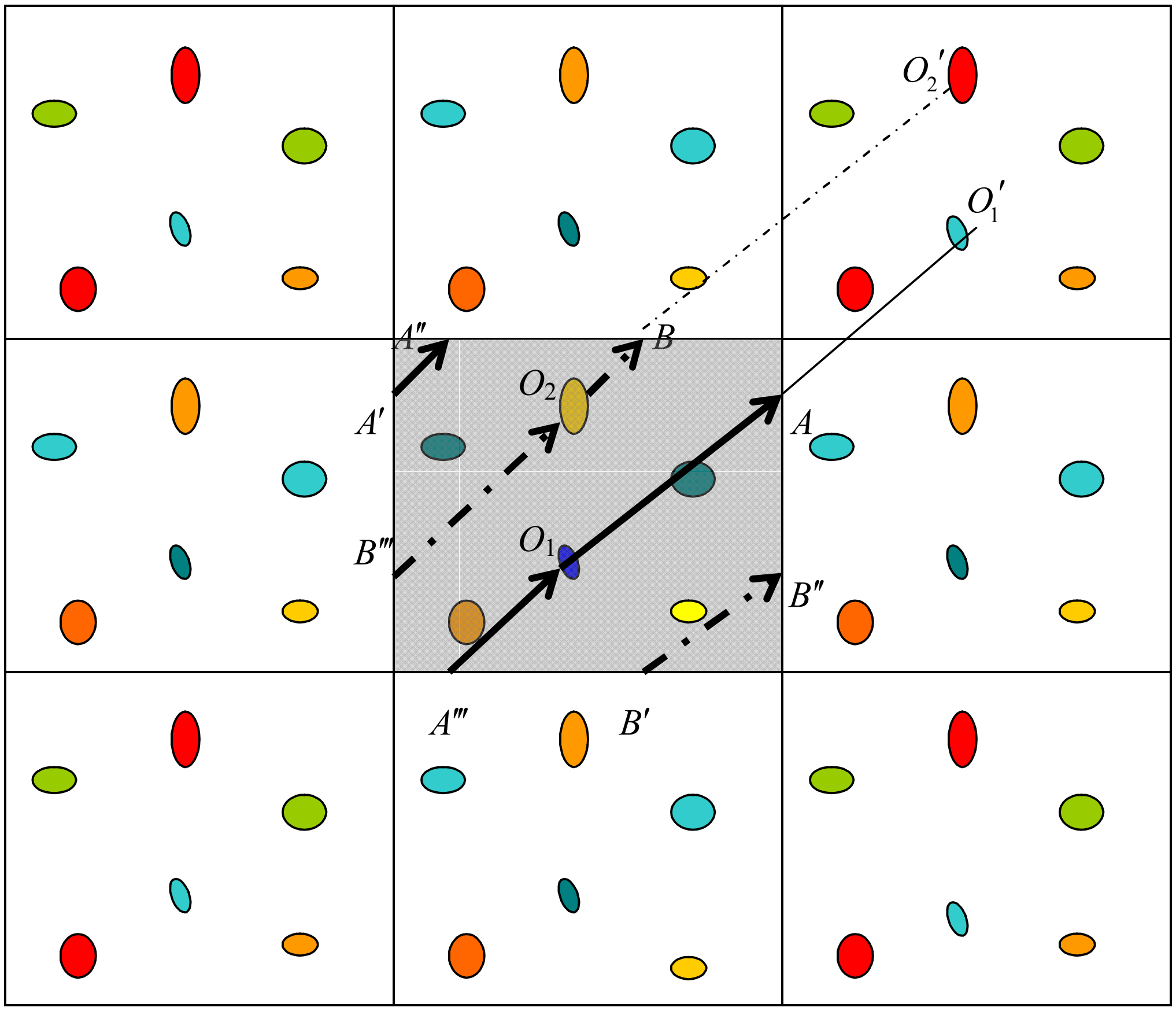

**Fig. 1.** See text for description of Figs. 1-4. Colors give a rough indication of redshift, assuming that all images are observable by $O_1$ or $O_2$.

which is a 2-dimensional version of $T^3$ as comoving spatial section of an Einstein-de Sitter cosmology. The nine rectangles in the figure are part of the tessellation generated in $E^2$ (the Euclidean plane) by the group $\Gamma$ of finite translations. The shaded area is the fundamental polyhedron *P*. Assume coordinates are applied to the figure, with origin at the center, *x*-axis in the horizontal direction, and *y*-axis in the vertical direction. The rectangles' edges have length *a* (horizontal) and *b*. Then $\Gamma$ is generated by $\gamma_1$: $(x,y) \to (x+a,y)$, and $\gamma_2$: $(x,y) \to (x,y+b)$. We take $O_1$ and $O_2$ as the positions of two observers, and $\gamma = \gamma_1\gamma_2$. Then loop $(O_1,\gamma)$ is represented on $\bar{P}$ by the union $O_1A \cup A'A'' \cup A'''O_1$, with $A \equiv A'$, $A'' \equiv A'''$; and $(O_2,\gamma)$ by $O_2B \cup B'B'' \cup B'''O_2$, with $B \equiv B'$, $B'' \equiv B'''$. These loops are the projections of their lifts $(O_1, O_1' = \gamma O_1)$ and $(O_2, O_2' = \gamma O_2)$ into $E^2$, hence $L(O_1,\gamma) = L(O_2,\gamma)$. Note that this figure can be thought of as a scaled representation, since the metric on $T^2$ is Euclidean.

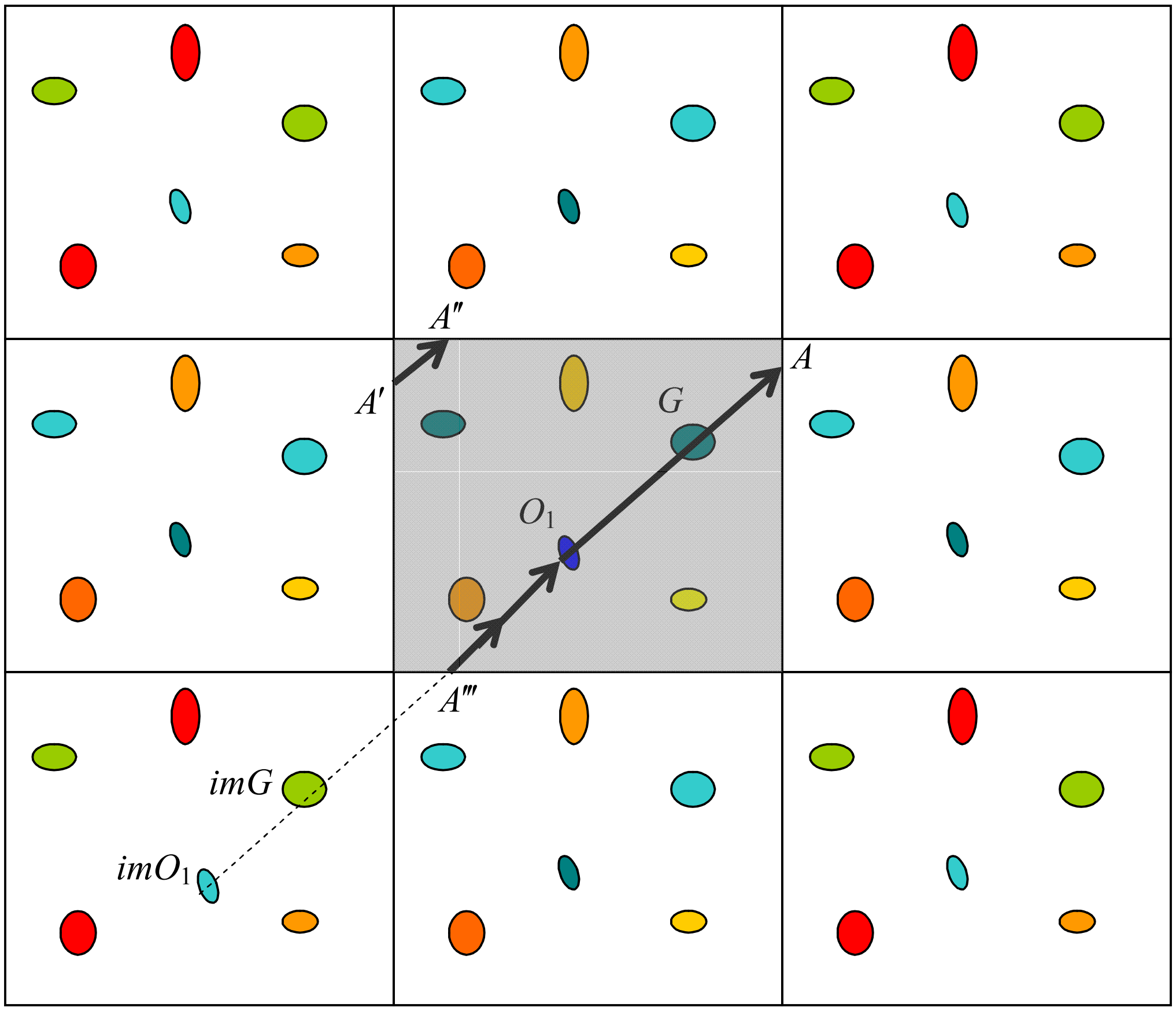

**Fig. 2**

*Theorem 3.2.* In a homogeneous space form $\Sigma$ the loops $(x,\gamma)$, $(x',\gamma)$, for any $x, x' \in \Sigma$, $\lambda \in \Gamma$, have the same loop angle.

*Proof.* If $\Sigma$ is homogeneous with respect to group of isometries $G$, then there is $g \in G$, such that $x' = gx$. And $G$ is the *centralizer* of $\Gamma$ in the group $\tilde{G}$ of isometries of $\tilde{\Sigma}$, i.e., $G = \{g \in \tilde{G}; \gamma g = g\gamma\}$; see Ref. 1, p.11. Then $g(x,\gamma) = \pi[g(\tilde{x},\gamma\tilde{x})] = \pi(g\tilde{x}, g\gamma\tilde{x}) = \pi(g\tilde{x}, \gamma g\tilde{x}) = (gx,\gamma) = (x',\gamma)$, that is, $(x',\gamma)$ is just the image of $(x,\gamma)$ under the action of $g$. So $\mathbf{t}'(0) = g_*\mathbf{t}(0)$, $\mathbf{t}'(1) = g_*\mathbf{t}(1)$, where $g_*$ is the map of tensor spaces induced by $g$. Therefore (cf. Ref. 23, p.43), $\cos[\mathrm{angle}(x',\gamma)] = g_*\mathbf{h}[g_*\mathbf{t}(0), g_*\mathbf{t}(1)] = \mathbf{h}[\mathbf{t}(0),\mathbf{t}(1)] = \cos[\mathrm{angle}(x,\gamma)]$. Hence $\mathrm{angle}(x',\gamma) = \mathrm{angle}(x,\gamma)$. ■

Fig. 1 also illustrates Theorem 3.2, in the trivial case of $T^2$, where all geodesic loops have null angle.

Observationally, the existence of a geodesic loop $(x,\gamma)$ implies that an astronomer at point $x$ might detect an old image of her of his own place in the universe, in the direction opposite $\mathbf{t}(1)$, at comoving distance $L(x,\gamma)$. (As stated above, a null geodesic of spacetime projects onto a geodesic of comoving space.) Direction $\mathbf{t}(0)$ could only be known theoretically in this case, since this is the *emission* direction of the ray, which was not observed. Another possibility is that a source be located at some point $z \in (x,\gamma)$, $z \neq x$; then the source's radiation might produce two images, the one located at $\tilde{z}$ in the direction $\mathbf{t}(0)$, the other at $\gamma\tilde{z}$ in the direction $-\mathbf{t}(1)$. Examples of these properties are given in Fig. 2, where $x = \tilde{x} = O_1$, $\gamma\tilde{x} = imO_1$; $z = \tilde{z} = G$, $\gamma\tilde{z} = imG$.

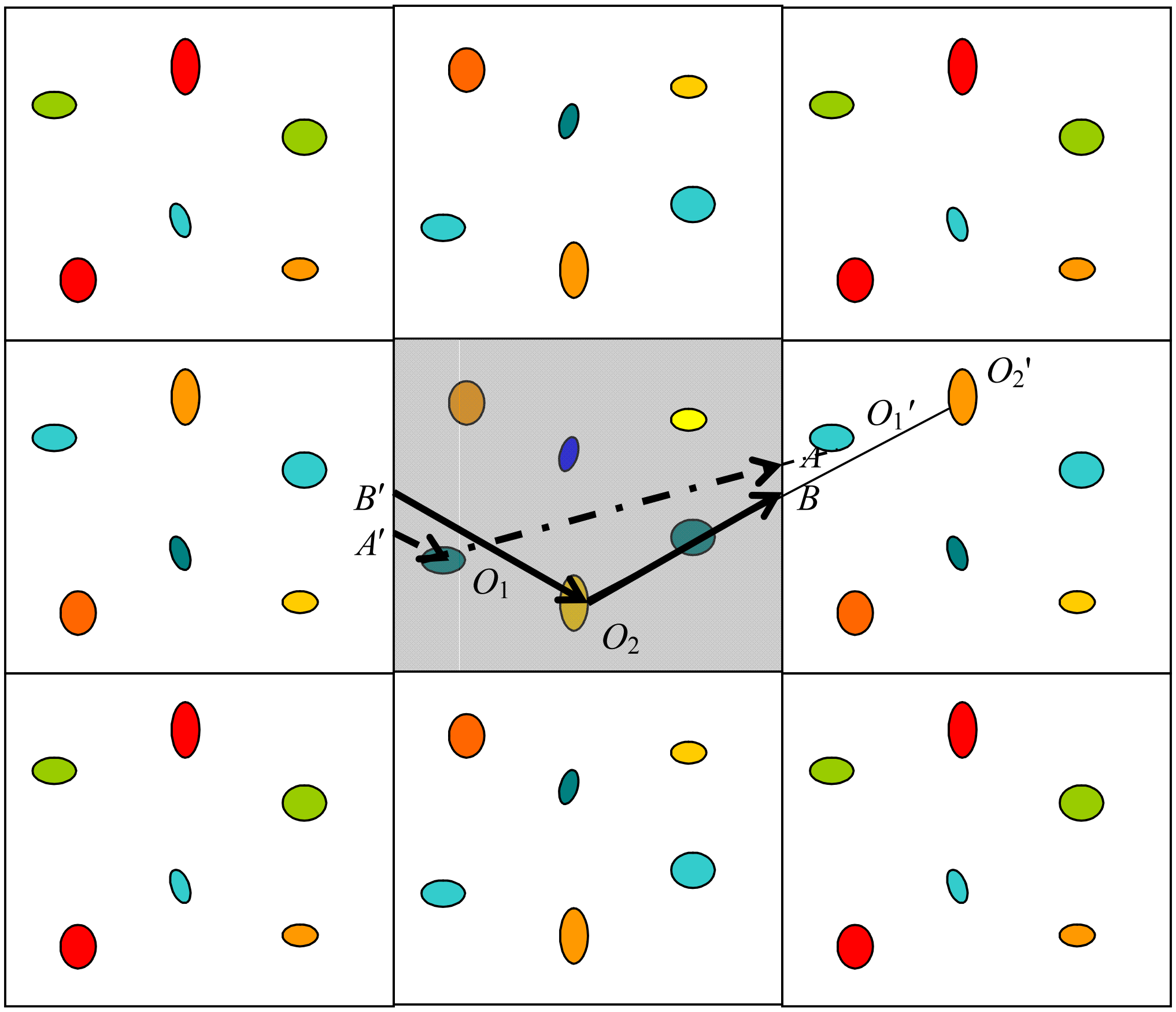

**Fig. 3**

To verify inhomogeneity of a space we only need find a violation of the consequences of Theorems 3.1 and 3.2. Fig. 3 shows a violation of the former, in the case of a 2-dimensional space with Euclidean metric and the topology of the Klein bottle. (As a surface the Klein bottle is non-orientable, but it may be viewed as a section of an orientable 3-space, namely the second of the flat space forms listed in Ref. 1.) In this case the fundamental region is the same as for $T^2$ (see above), but the generators are $\gamma_1$ : $(x,y) \to (x+a,-y)$, and $\gamma_2$: $(x,y) \to$ $(x,y+b)$. For $\gamma = \gamma_1$, Fig. 3 shows that $D(O_1,\gamma O_1) \neq D(O_2,\gamma O_2)$, hence $L(O_1,\gamma) \neq L(O_2,\gamma)$. Fig. 4 shows, for the same space as in Fig. 3, a violation of Theorem 3.2: angle$(O_1,\gamma) \neq$ angle$(O_2,\gamma)$.

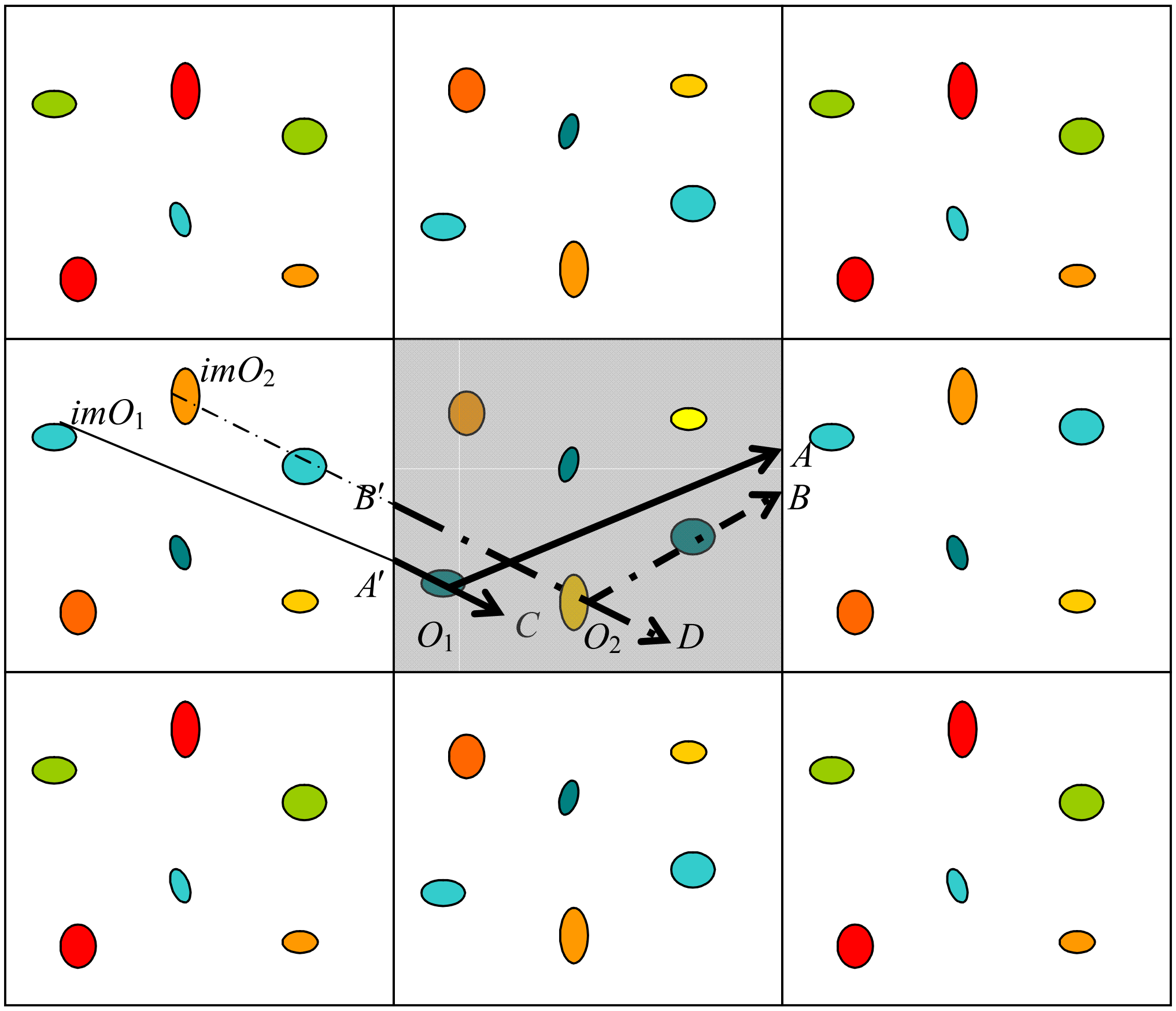

**Fig. 4**

In practice it should be sufficient to examine generator loops $(p,\gamma)$, $(q,\gamma)$ for suitable pairs $p$, $q$ $\in \Sigma$. This was done with a model having closed spatial sections of negative curvature [22]. From two possible positions of our galaxy in the fundamental polyhedron (and hence in $\Sigma$), two sets of early images of the Galaxy were obtained from the generator loops. The distances of corresponding potential images in the two sets are different, hence $\Sigma$ does not satisfy Theorem 3.1 and so is inhomogeneous. This is as expected, since (Ref. 14, Theorem 2.7.1) no hyperbolic closed space is homogeneous. But the defining property of inhomogeneity, namely lack of a global transitive group of isometries, is too abstract an observational viewpoint; while the above result on sets of potential self-images of the Galaxy gives a graphical meaning to lack of homogeneity.

### 3.3. The homogeneity of the space of images

Large scale cosmic observations support (as an approximation) the theoretical assumption of spatial isotropy in FRW models, which have simply connected spatial topologies. Also, the Copernican principle leads to the homogeneity postulate for these models. In the case of a model with multiply connected space sections, these assumptions are generally no true. The next theorem elaborates on an argument, previously presented in sketchy form [25], to show how this difficulty is circumvented.

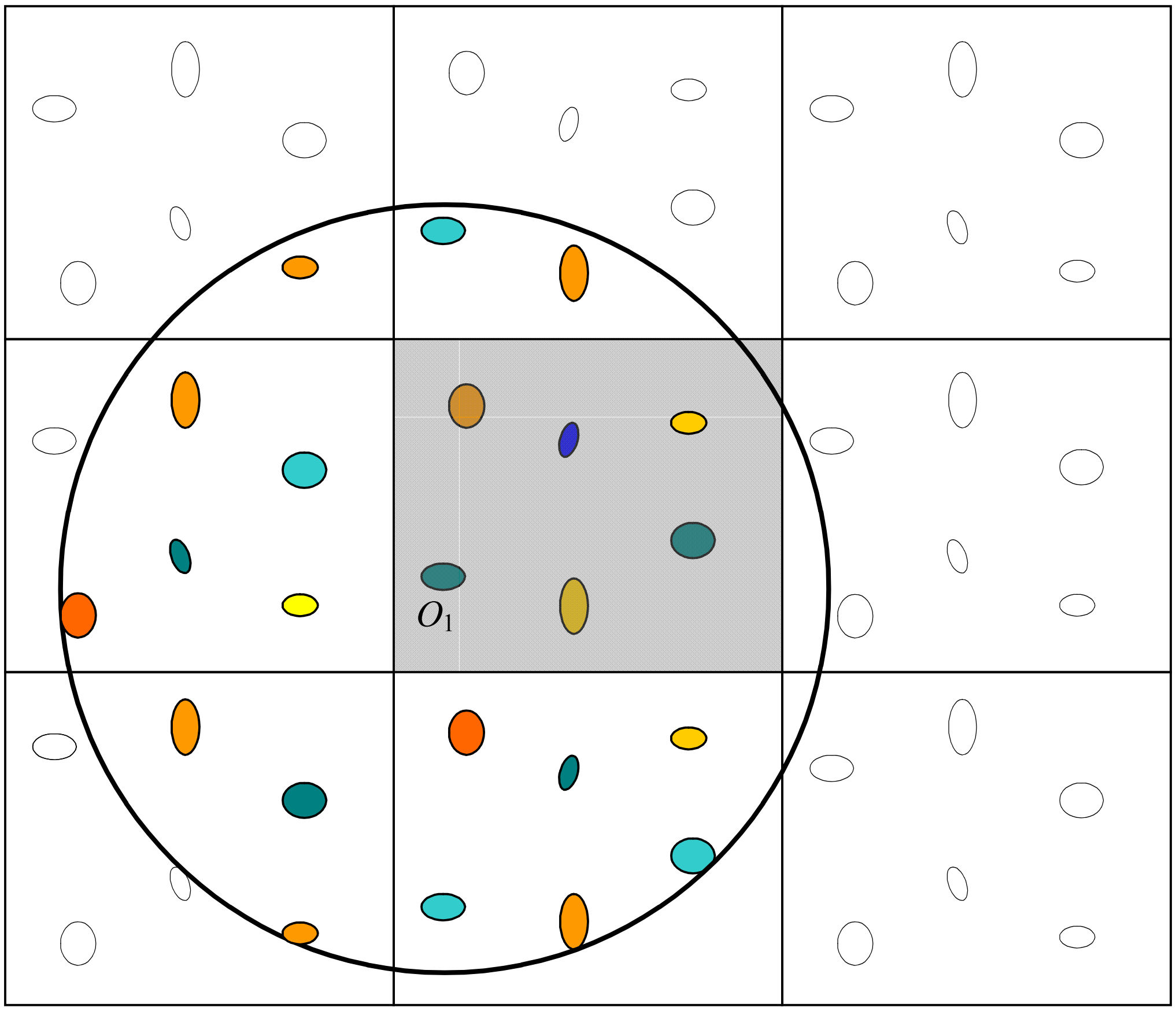

**Fig. 5.** Observable universe for observer at $O_1$, showing homogeneity and isotropy of the distribution of images. Unobservable images are left blank.

*Theorem 3.3*. If the density of radiation sources in $\Sigma$ is uniform, so is the density of potential images of these sources in $\widetilde{\Sigma}$, the universal covering space of $\Sigma$.

*Proof*. Since $\Sigma \cong \overline{P}$ (cf. Section 3.1), and $\Gamma$ is a group of deck transformations (see Ref. 18, p. 102), it follows that $\widetilde{\Sigma}$ is tessellated by the copies $\gamma(P)$ or $\gamma P$ of the fundamental polyhedron $P$ of $\Sigma$. That is, $\widetilde{\Sigma}$ is covered by these cells, and the interiors of different cells do not overlap; in symbols: $\widetilde{\Sigma} = \mathrm{U}_{\gamma\in\Gamma}(\gamma P)$, and if $\gamma_1 \neq \gamma_2$ then $\mathrm{int}(\gamma_1 P) \cap \mathrm{int}(\gamma_2 P) = \varnothing$. Let $p$ be the observer's position and $q$ the position of a source $S$ in $P$. For each $\gamma \in \Gamma$, the geodesic segment $(\gamma q, p)$ in $\tilde{\Sigma}$ is projected onto a different geodesic segment from $q$ to $p$ in $\Sigma$. Since the latter are radiation paths in 3-dimensional space, they may produce different images of the source, which are seen at $p$ as apparently coming from $\gamma q$. Hence each cell $\gamma(P)$ contains a potential image of $S$. (They are potential in the same sense that a ray emitted by an actual source at $\gamma q$ in the usual models may or may not reach us in our cosmic moment.) Since $\gamma$ is an isometry of $\tilde{\Sigma}$, the pattern of image points $\gamma q$ in $\gamma(P)$ is the same as the pattern of source points $q$ in $P$, because isometries preserve distances and angles. Therefore, if the density of sources $q$ in $P$ and $\Sigma$ is uniform, so is the density of image points $\gamma q$ in each $\gamma(P)$, and hence in $\tilde{\Sigma}$.
■

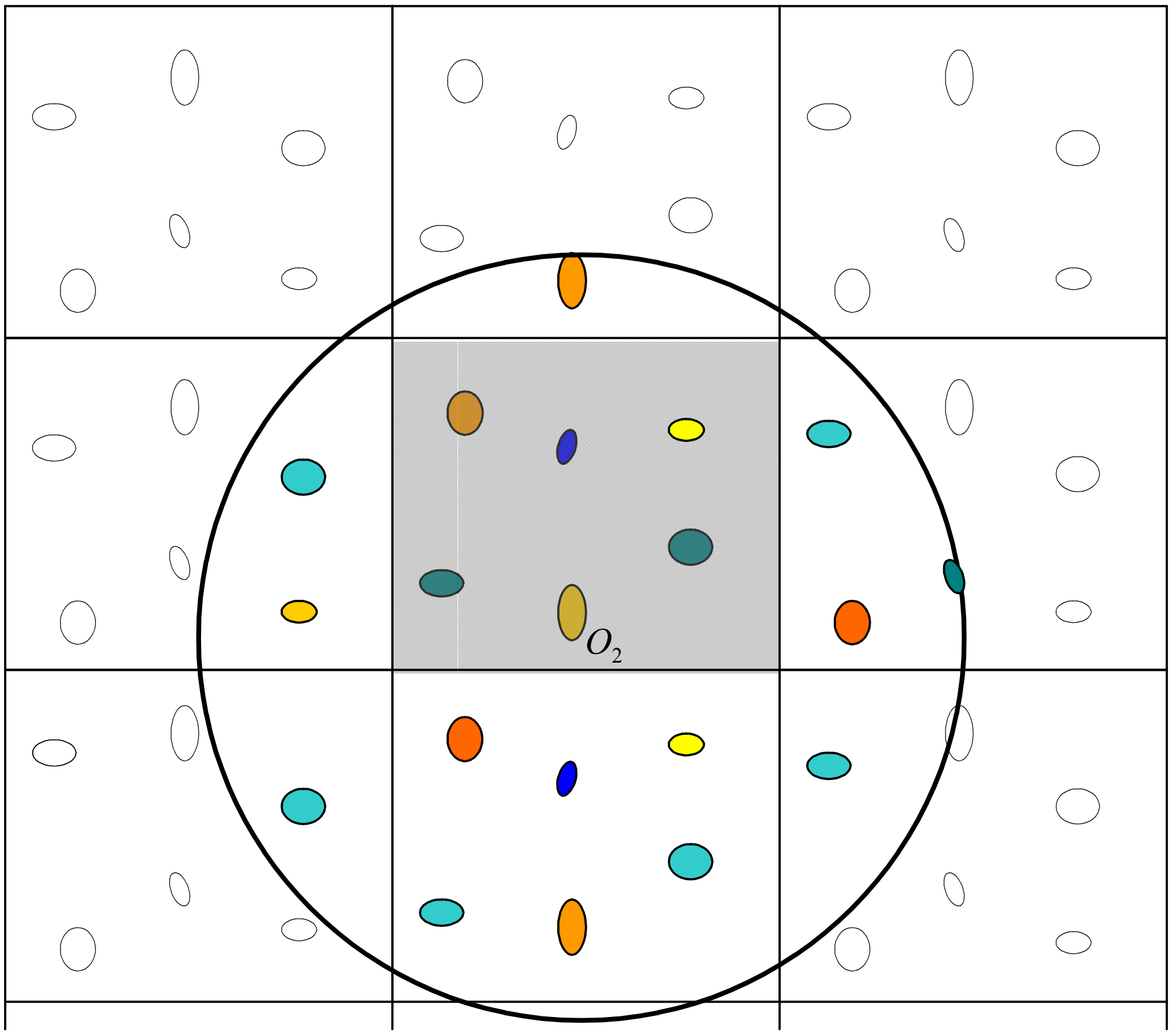

**Fig. 6.** As in Fig. 5, now for observer at $O_2$.

The apparent homogeneity implies an apparent isotropy for any observer. See Fig. 1 in Ref. 13, for an illustration of this last property in terms of the Euclidean torus $T^2$. Figs. 5 and 6 are crude illustrations of the homogeneity and isotropy of the distribution of images within the particle horizons (the circles) of two observers $O_1$ and $O_2$, respectively, for the same space as in Figs. 3 and 4.

## 4. FINAL COMMENTS

Unfortunately, present data do not allow us to decide whether cosmic space is simply or multiply connected – see Ref. 26, for example. However, as more and better observations are made, we may hope that sooner or later it may be possible to establish whether there exist cosmic structures showing repetitive or evolutionary patterns, and whether such patterns can be interpreted as the effect of a cosmic nontrivial topology.

Elementary particle theorists may object that for spaces of complex topology it may not be possible to have well defined spinor fields (see Ref. 17, Ch. 1) that would describe the abundance of neutrinos and other particles in the universe. From the discussion by Geroch [28] we know that our spacetimes do admit spinor structure, since their topology is of the form $R \times S(t)$, with $S(t)$ spacelike and orientable. This structure is not unique for multiply connected spatial sections, but we can choose one of the two alternatives as part of the *definition* of spinors – cf. Ref. 27. Anyway, mathematical physicists are looking for alternatives to the usual SL(2,**C**) spinor structure formalism – see Refs. 29, 30, and references therein.

## ACKNOWLEDGEMENTS

I am grateful to Sandro Silva e Costa for pointing out to me some mistakes in the original Table III; to Ruben Aldrovandi for a discussion on an early draft of Section 3; to Alessandro Figà-Talamanca, who provided me with a copy of Bianchi's centenary monograph [3]; and to many colleagues and students, whose questioning made me rethink and improve on a first version of this paper.

[4] After publication of this paper, an English translation of Bianchi's article, by Robert Jantzen, appeared as a 'golden oldie' in *Gen. Rel. Grav.* 33, 2171 (2001).

**References added in this arXiv version**

[5] Now *Brazilian Journal of Physics.*